\documentstyle[12pt,psfig,here]{article}
\newcommand{\be}{\begin{equation}}
\newcommand{\ee}{\end{equation}}
\newcommand{\bea}{\begin{eqnarray}}
\newcommand{\eea}{\end{eqnarray}}
\newcommand{\bean}{\begin{eqnarray*}}
\newcommand{\eean}{\end{eqnarray*}}
\newcommand{\ba}{\begin{array}}
\newcommand{\ea}{\end{array}}

\newcommand{\norsl}{\normalsize\sl}
\newcommand{\norsc}{\normalsize\sc}

\begin{document}

\begin{titlepage}
\title{Electroweak Sudakov at two loop level}

\author{
\norsc  M. HORI\thanks{JSPS Research Fellow}\ ,
        H. KAWAMURA\thanks{JSPS Research Fellow}\  and J. KODAIRA\\
\norsl  Dept. of Physics, Hiroshima University\\
\norsl  Higashi-Hiroshima 739-8526, JAPAN\\}

\date{}
\maketitle

\vspace*{1.5cm}

\begin{abstract}
{\normalsize
\noindent
We investigate the Sudakov double logarithmic corrections
to the form factor of fermion in the SU(2)$\otimes$U(1) electroweak theory.
We adopt the familiar Feynman gauge and present explicit calculations
at the two loop level.
We show that the leading logarithmic corrections coming from the infrared 
singularities are consistent with the \lq\lq postulated\rq\rq\  exponentiated
electroweak Sudakov form factor.
The similarities and differences in the \lq\lq soft\rq\rq\ 
physics between the electroweak theory
and the unbroken non-abelian gauge theory (QCD) will
be clarified.}

\end{abstract}

\begin{picture}(5,2)(-330,-485)
\put(2.3,-110){HUPD-0003}
\end{picture}

\vspace{4cm}
\leftline{\hspace{1.2cm}hep-ph/0007329}
\leftline{\hspace{1.2cm}July 2000}
\thispagestyle{empty}

\end{titlepage}
\setcounter{page}{1}
\baselineskip 18pt 
\section{Introduction}
Recently the high energy behavior of the Standard Model (SM) electroweak
theory receives much attention from both theoretical and
phenomenological viewpoints.
At future high energy colliders, the total energy is much bigger than
the masses of the SM gauge bosons and large double logarithmic (DL)
corrections originating from the infrared behavior of the gauge
theory~\cite{SU} can not be neglected for the exclusive~\cite{CC1}
and also for the inclusive~\cite{CCC} processes.

The infrared behavior of the gauge theory has been one of the main
subjects of the particle physics and many investigations have been
made mainly for the QED and the unbroken non-abelian gauge theory (QCD).
For the form factor of fermion in QED and QCD, it is known that
the leading singularities can be exponentiated resulting
in the Sudakov form factor~\cite{JCC}.
In the case of the SM electroweak theory,
the situation is much more complicated than QCD in two aspects.
The first is that the symmetry is spontaneously broken.
Secondly, the pattern of the symmetry breaking is not diagonal,
namely a particular combination of the direct product of the
gauge group SU(2)$\otimes$U(1) is survived as an unbroken
gauge group ${\rm U(1)}_{\rm em}$.
These mean that we must carefully examine the non-abelian structure
and the mixing effect which leads to \lq\lq mass gap\rq\rq\ between
the gauge bosons in the electroweak theory.
Therefore it is never trivial that the infrared singularities
which appear in the form factor
can be exponentiated as in QED and QCD.
Several authors have addressed this
problem~\cite{CC2,KP,KPS,FLMM,M1,BW}.
The authors in Ref.\cite{CC2} used the formalism which has been
developed in QCD~\cite{C}.
The infrared evolution equation~\cite{KL} has been adopted in
Refs.\cite{FLMM,M1}.
The explicit two loop calculations have been done by the authors in
Refs.\cite{KP,BW} in the Coulomb or axial gauge.
Unfortunately the situation is still somehow ambiguous
concerning the possibility of the exponentiation of the infrared
singularity in the electroweak theory.

Recently, a part of the mixing effect (mass gap effect) mentioned above
is investigated by Melles~\cite{M2} in the Feynman gauge.
In this paper, we extend the analysis by Melles to the 
general case which includes also the information on the non-abelian
structure of the electroweak theory.
Since there are many investigations in the Feynman gauge for QCD,
we believe that our explicit two loop calculation
of fermion's form factor in the Feynman gauge is useful and instructive.
Furthermore we will be able to clarify the similarities and
differences in the \lq\lq soft\rq\rq\ 
physics between the electroweak theory and QCD.
The process we consider is the fermion pair production from the
SU(2)$\otimes$U(1) singlet source.
To the accuracy of the DL corrections,
the chirality of the fermion is conserved and we can discuss the
left and right handed fermion separately.
Therefore we consider, in this paper, the left handed fermion
and the right handed antifermion production\footnote{
The case of the right handed fermion and the left handed antifermion
production is discussed by Melles~\cite{M2}}.
The masses of the $W$ and $Z$ bosons will be approximated to be equal
$M_W \sim M_Z \equiv M$.
We give a fictitious small mass $\lambda$ to the photon to regularize
the \lq\lq real\rq\rq\  infrared divergence and the fermion is
assumed to be massless.
We assume the situation, $s \gg M^2 \gg \lambda^2$ with
$s$ the total energy of the produced fermions.

\section{One loop DL contribution}
To fix our convention and the calculational framework,
we present the Feynman rule (in the Feynman gauge)
and the one loop calculation.
The Feynman rules for the gauge boson propagators
and the fermion gauge boson couplings read~\footnote{Since we assume
fermions to be massless, the ghost and Higgs particles do not
contribute.},
\[   \gamma \quad : \quad \frac{-i g_{\mu\nu}}{q^2 - \lambda^2}
      \qquad , \qquad W^{\pm},Z \quad : \quad 
             \frac{-i g_{\mu\nu}}{q^2 - M^2} \ ,\]
and
\bea
    \gamma f f &:& i e Q ( \gamma_{\mu} \omega_- + \gamma_{\mu}
               \omega_+ )\ ,\nonumber\\
    W^{\pm} f f &:& \frac{i g}{\sqrt{2}}
           ( T^1 \pm i T^2 ) \gamma_{\mu} \omega_-
            \ , \label{f-rule}\\
    Z f f &:& \frac{i g}{\cos \theta_W}
              \left[ ( T^3 - \sin^2 \theta_W Q ) \gamma_{\mu} \omega_-
                - \sin^2 \theta_W Q \gamma_{\mu} \omega_+ \right]
           \ ,\nonumber
\eea
where $T^a (a = 1,2,3)$ is the SU(2) generator, $Q$ is the charge
of fermion given by $Q = T^3 + Y$ with $Y$ the heypercharge
and $\theta_W$ is the Weinberg angle.
$\omega_{\pm} \equiv \frac{1 \pm \gamma_5}{2}$ is the projection operator.
The coupling constants of the SU(2) and U(1) gauge groups
are $g$ and $g' = g \tan \theta_W$ respectively and
the electric charge $e ( > 0)$ is related to $g$ as $e = g \sin
\theta_W$.
As explained in the Introduction,
we concentrate on only the left chiral part
in Eq.(\ref{f-rule}) in what follows.

Let us present the group factor of SU(2)$\otimes$U(1) and
the \lq\lq kinematical\rq\rq\ factor from loop integration separately. 
The group factors become:
\bean
    \gamma\ {\rm exchange}\ &:& e^2 Q^2 \ ,\\
     W \ {\rm exchange}\ &:& g^2 \sum_{a= 1,2} T^a T^a \ ,\\
     Z \ {\rm exchange}\ &:& \frac{g^2}{\cos^2 \theta_W}
            ( T^3 - \sin^2 \theta_W Q )( T^3 - \sin^2 \theta_W Q )\\
                & & \qquad = g^2 T^3 T^3 + g'^2 Y^2 - e^2 Q^2 \ .
\eean
The loop integrations in which the weak bosons and photon are exchanged
produce the double logarithmic corrections,
\[   - \frac{1}{16 \pi^2} \ln^2 \frac{s}{M^2}
        \quad , \quad - \frac{1}{16 \pi^2} \ln^2 \frac{s}{\lambda^2}\ ,\]
respectively.
Therefore, the final result up to the one loop level for the fermion pair
production reads,
\[    \Gamma^{(1)} =  1 - 
           \frac{1}{16 \pi^2} \left( g^2 C_2 (R) + g'^2 Y^2 - e^2 Q^2 \right)
              \ln^2 \frac{s}{M^2} - \frac{1}{16 \pi^2} e^2 Q^2
                  \ln^2 \frac{s}{\lambda^2} \ , \]
where $C_2 (R)$ is the SU(2) Casimir operator for the fundamental
representation.

\section{Two loop DL contribution}
We classify the two loop diagrams into three groups.
The first group is composed from the ladder and crossed ladder
diagrams shown in Fig.1.
The second includes the diagrams which contain the triple gauge
boson coupling shown in Fig.2.
The symmetric graphs are not shown in Figs.1 and 2.
The diagrams with the vertex correction and the self energy insertion
make the third group (graphs are not shown).

For the evaluation of diagrams in the second group,
it is known that the application of the pinch technique~\cite{CP}
to the triple gauge boson coupling makes the situation simple~\cite{FFT}.
Omitting the coupling constant, the triple gauge boson
vertex will be decomposed into two parts.
\bean
   \Gamma_{\alpha\mu\nu} (k , q)
      &=& (k - q)_{\nu} g_{\alpha\mu} + (2q + k)_{\alpha}
           g_{\mu\nu} - (q + 2k)_{\mu} g_{\nu\alpha} \\
      &=& \Gamma^P_{\alpha\mu\nu}
            + \Gamma^F_{\alpha\mu\nu} \ ,
\eean
with
\bea
  \Gamma^P_{\alpha\mu\nu} &\equiv&
           - (k + q)_{\nu} g_{\alpha\mu} - q_{\mu} g_{\nu\alpha} \ ,
           \label{pinchpart}\\
  \Gamma^F_{\alpha\mu\nu} &\equiv&
           (2 q + k)_{\alpha} g_{\mu\nu} + 2 k_{\nu} g_{\alpha\mu}
           - 2 k_{\mu} g_{\nu\alpha} \ .\label{nwt}
\eea
%
\begin{figure}[H]
\begin{center}
\begin{tabular}{ccc}
\leavevmode\psfig{file=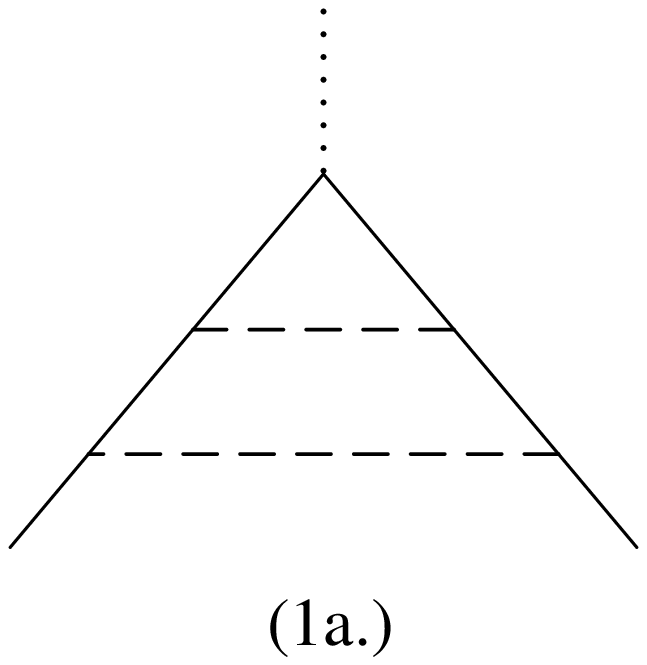,width=3cm} \quad  &
\leavevmode\psfig{file=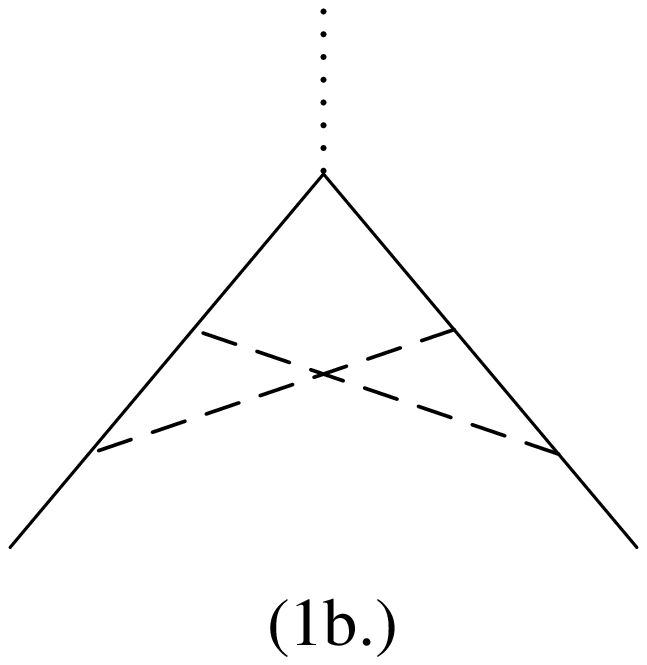,width=3cm}  & \\
& & \\
\leavevmode\psfig{file=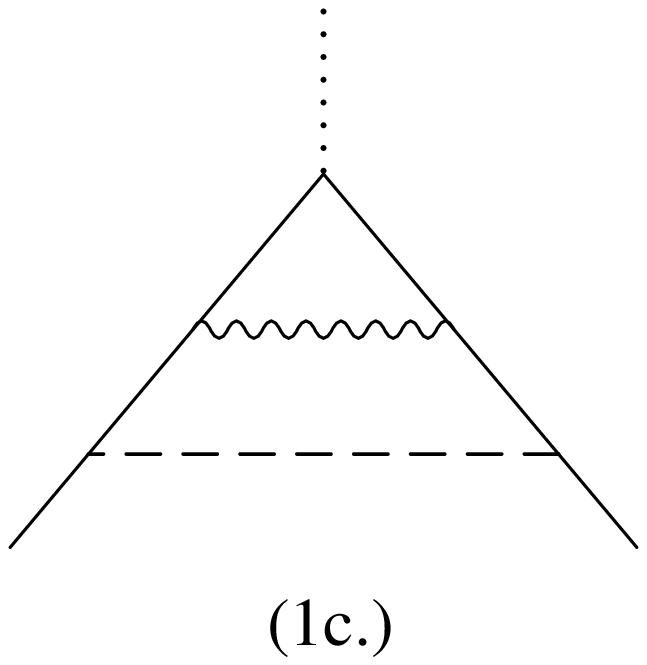,width=3cm} \quad  &
\leavevmode\psfig{file=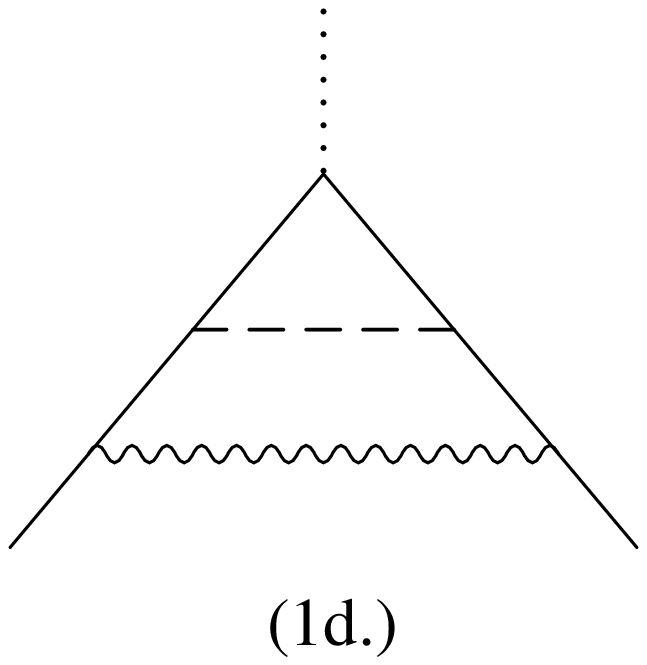,width=3cm} \quad  &
\leavevmode\psfig{file=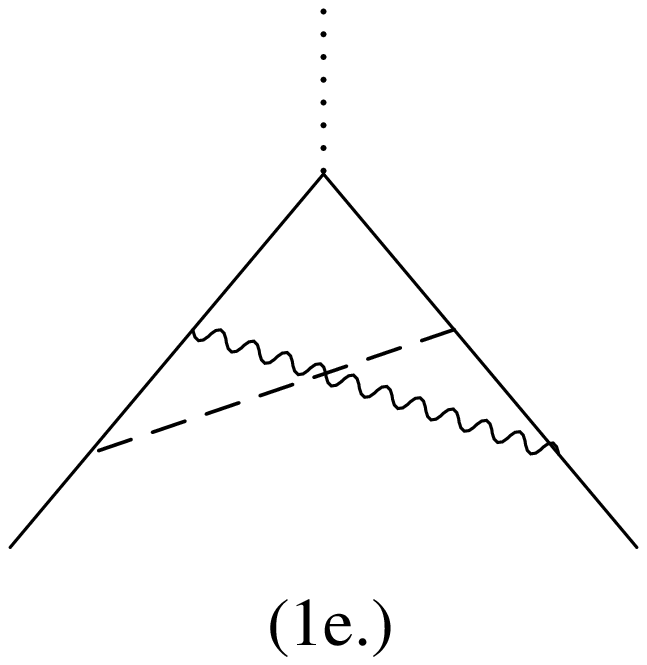,width=3cm}  \\
& & \\
\leavevmode\psfig{file=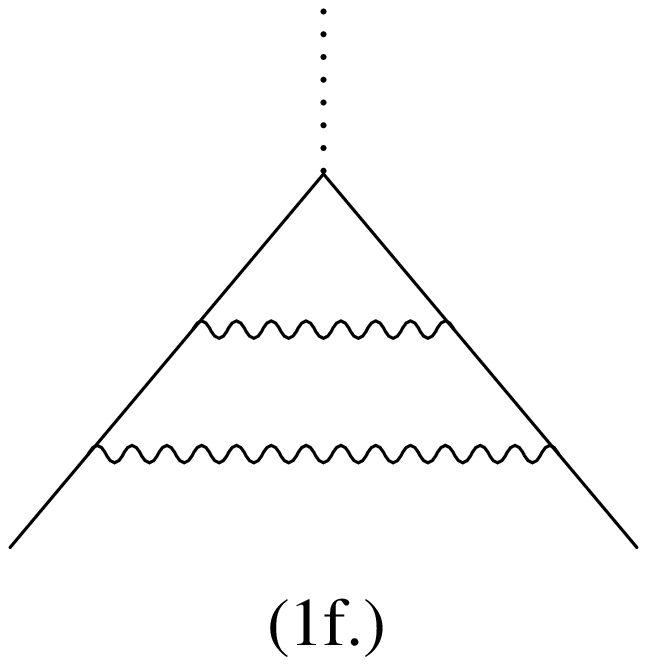,width=3cm} \quad &
\leavevmode\psfig{file=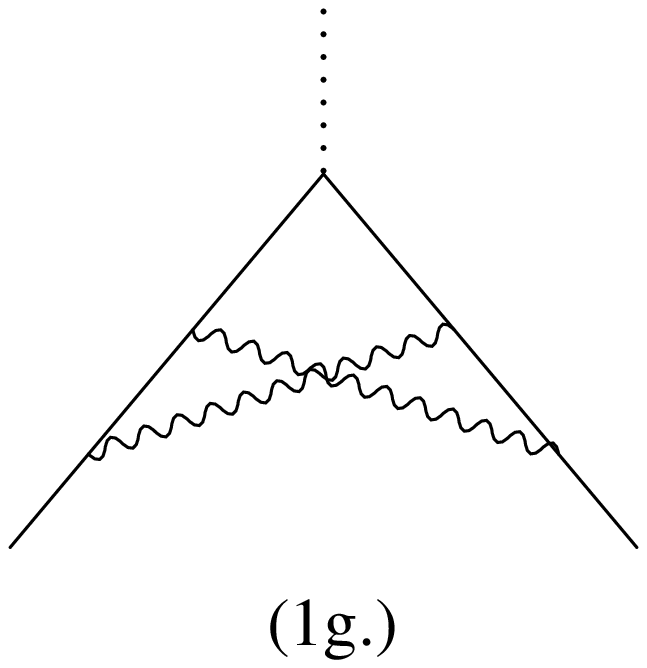,width=3cm} & 
\end{tabular}
\caption{The ladder and crossed ladder diagrams.
The dashed (wavy) line represents the photon ($W$ and/or $Z$)
with the mass $\lambda$ ($M$).}
\label{diag1}
\end{center}
\end{figure}
\vspace{-0.3cm}
\begin{figure}[H]
\begin{center}
\begin{tabular}{cccc}
\leavevmode\psfig{file=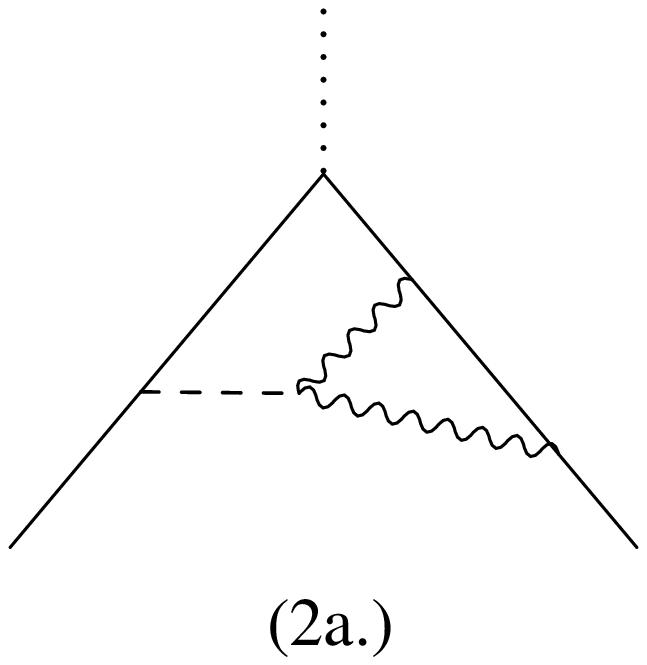,width=3cm}  &
\leavevmode\psfig{file=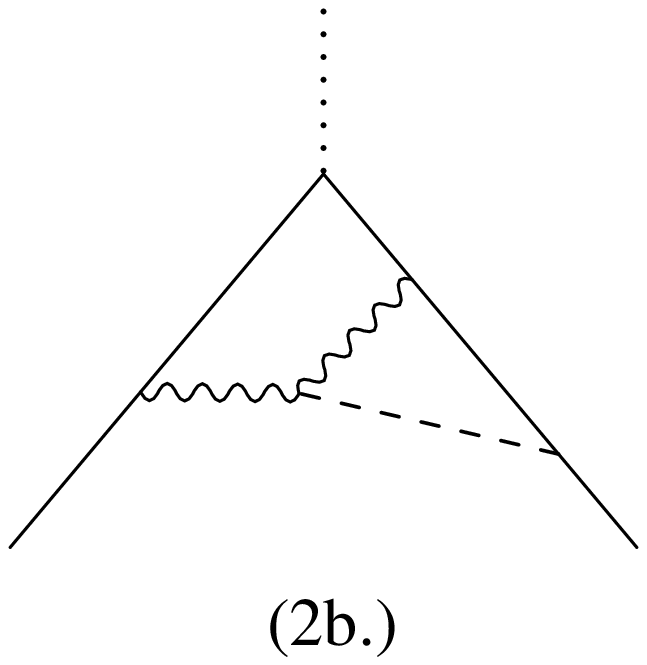,width=3cm}  &
\leavevmode\psfig{file=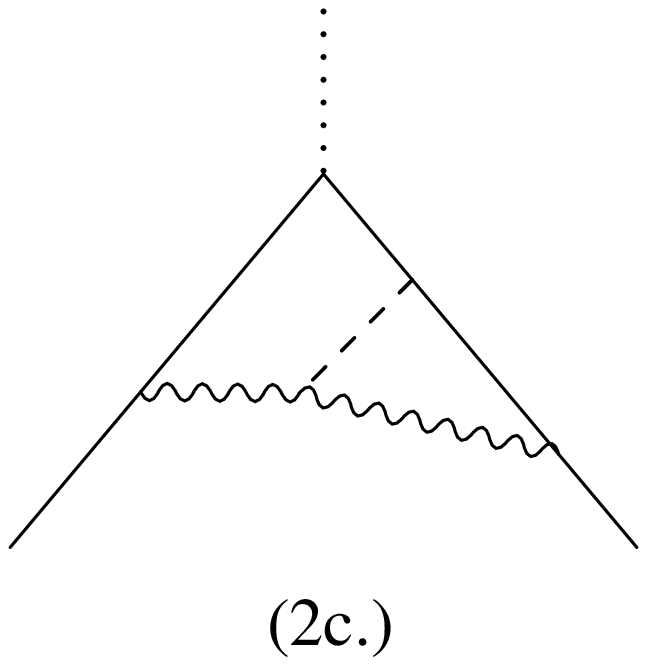,width=3cm}  &
\leavevmode\psfig{file=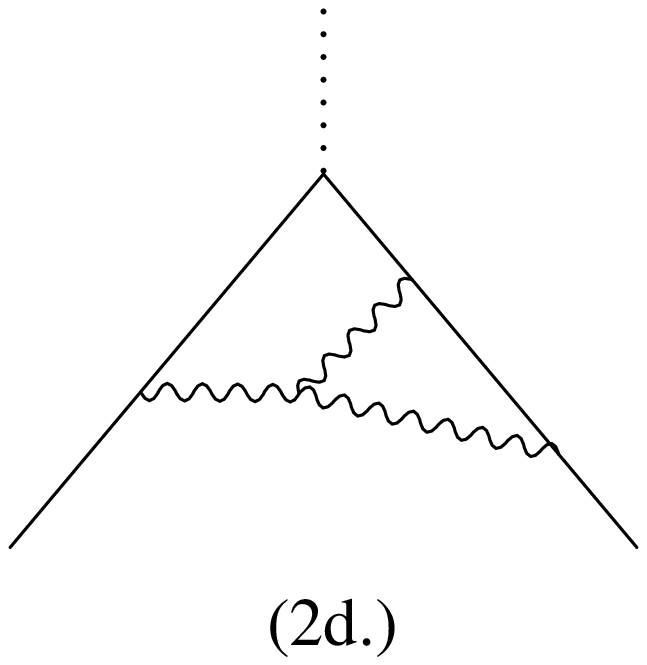,width=3cm}
\end{tabular}
\caption{The diagrams which have the triple point couplings.
The meaning of lines is the same as in Fig.\ref{diag1}.}
\label{diag2}
\end{center}
\end{figure}
\vspace{-0.3cm}
\noindent
Eq.(\ref{pinchpart}) gives rise to pinch parts when contracted
with $\gamma$ matrices and it is easily seen that the contributions from
this term are reduced to the effective diagrams shown in Figs.3b and 3c.
The contribution from Eq.(\ref{nwt}) is depicted in Fig.3a with
the blob vertex.
\begin{figure}[H]
\begin{center}
\begin{tabular}{ccc}
\leavevmode\psfig{file=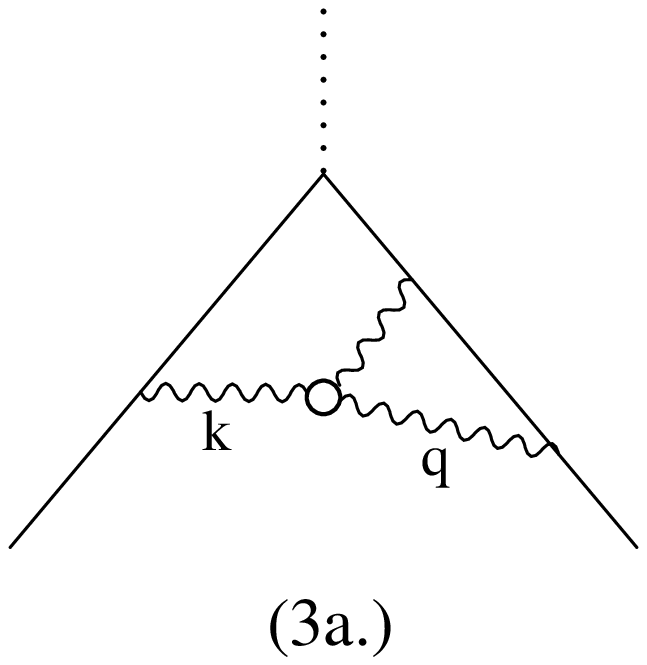,width=3.2cm} \quad &
\leavevmode\psfig{file=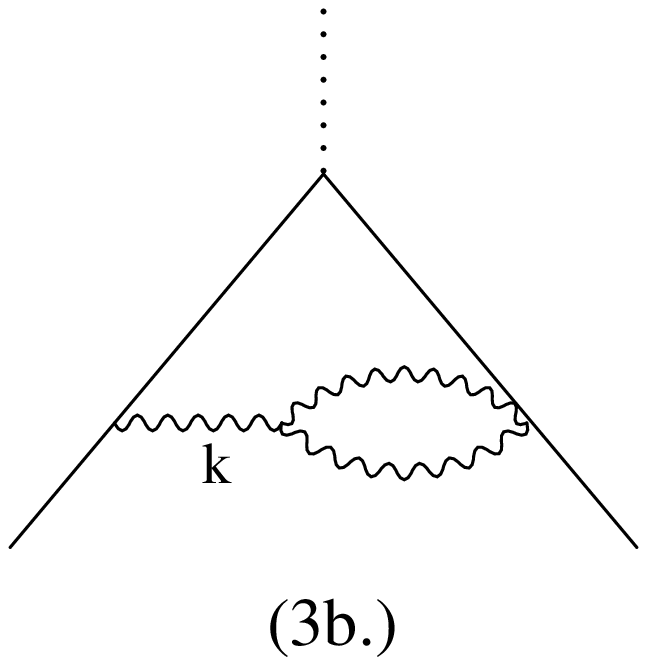,width=3.2cm} \quad &
\leavevmode\psfig{file=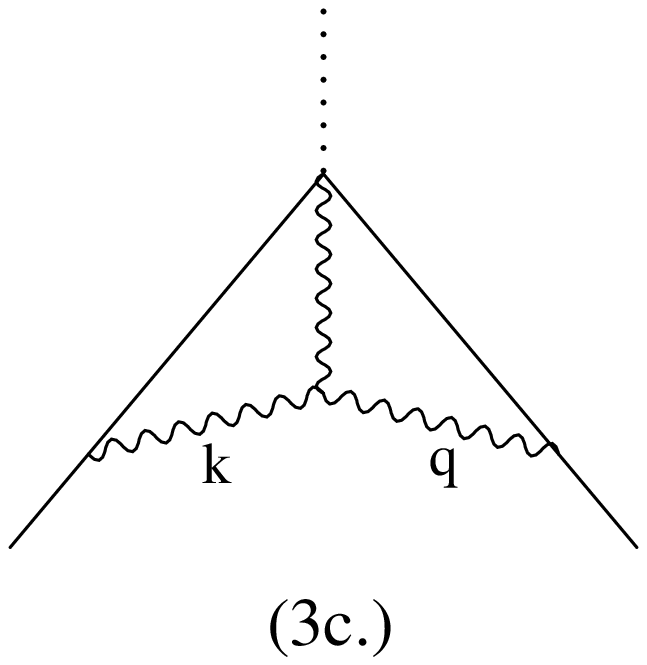,width=3.2cm}  
\end{tabular}
\caption{The reduced Feynman diagrams for Fig.\ref{diag2}.
The blob stands for the vertex $\Gamma^F$ defined in Eq.\ref{nwt}.
The wavy lines represent both the photon and $W , Z$.}
\label{diag3}
\end{center}
\end{figure}
\vspace{-0.3cm}
\noindent
By combining the contributions from Fig.3a (Fig.3b) with those from
the diagrams with the vertex correction (self energy insertion),
one will get the \lq\lq gauge invariant\rq\rq\  vertex (propagator).
The important fact for our purpose is that
these contributions are less singular at least by one power of $\ln s$
compared to diagrams Fig.1 and Fig.3c~\cite{FFT}.
Therefore, at the DL level, it is sufficient to calculate
diagrams Fig.1 and Fig.3c configurations of Fig.2.  
 
We present again the contributions from the group factor of
SU(2)$\otimes$U(1) and the \lq\lq kinematical\rq\rq\ factor
from loop integration separately because such
presentation is very useful to clarify the similarities and
differences between the electroweak theory and QCD.

The group factors for the ladder and crossed ladder diagrams in Fig.1 
become,
\bea
    (1a.) &=& (1b.) = ( e^2 Q^2 )^2 \label{gf1ab}\ ,\\
    (1c.) &=& e^2 Q^2 \left( g^2 C_2 (R) + g'^2 Y^2 - e^2 Q^2 \right)\ ,
             \label{gf1c}\\
    (1d.) &=& e^2 Q^2 \left( g^2 C_2 (R) + g'^2 Y^2 - e^2 Q^2 \right)
             - 2 g^2 e^2 Y T^3 \ ,\label{gf1d}\\
    (1e.) &=& 2 \times \left[ e^2 Q^2 \left( g^2 C_2 (R)
            + g'^2 Y^2 - e^2 Q^2 \right)
             -  g^2 e^2 Q T^3 \right]\ ,\label{gf1e}\\
    (1f.) &=& \left( g^2 C_2 (R) + g'^2 Y^2 - e^2 Q^2 \right)^2
                + 2 g^2 e^2 Y T^3 \ , \label{gf1f}\\
    (1g.) &=& \left( g^2 C_2 (R) + g'^2 Y^2 - e^2 Q^2 \right)^2
                - g^4 C_2 (R) + 2 g^2 e^2 Q T^3 \label{gf1g}\ .
\eea
Those for the diagrams having the triple gauge boson couplings (Fig.2)
read,
\bea
    (2a.) &=&  (2b.) = 2 \times \left[ g^2 e^2 Q T^3 \right]\ ,\label{gf2a}\\
    (2c.) &=& 2 \times \left[ - g^2 e^2 Y T^3 + g^2 e^2 T^3 T^3 \right]
              \label{gf2b}\ ,\\
    (2d.) &=& 2 \times \left[ g^4 C_2 (R) - 2 g^2 e^2 Q T^3
                + g^2 e^2 Y T^3 - g^2 e^2 T^3 T^3 \right]
                  \ . \quad\qquad \label{gf2c}
\eea
The factor $2$ in Eqs.(\ref{gf1e},\ref{gf2a},\ref{gf2b},\ref{gf2c})
comes from the symmetric diagram.
One can see that the structure of the group factors is
slightly more complicated than that of QCD.
In QCD, only the Casimir operator $C_2(R)^2$
appears in the ladder diagram.
The crossed ladder diagram is proportinal
to $C_2(R)^2 - \frac{1}{2} C_2(R) C_2(G)$ ($C_2(G)$ is the
Casimir for the adjoint representation) and the second
term is a non-exponentiating term.
However it is known that this term is canceled by the contribution
from Fig.2.
On the other hand, the situation is different in the electroweak theory.   
The non-abelian nature of 
SU(2) and the mixing effect between SU(2) and U(1)
lead to new contributions compared to QCD.
Since the propagation of the electroweak bosons
is not proportional to the Casimir operator,
there are non-exponentiating terms even in the ladder diagrams.
See Eqs.(\ref{gf1d},\ref{gf1f}) (Figs.1d, 1f).
The crossed ladder diagrams and Fig.2
receive the contributions which originate from both the mixing
effect and the non-abelian nature of SU(2).
The latter is the same as in QCD.

To evaluate the loop integrals, we follow the method explained
in Ref.\cite{M2} for the ladder and crossed ladder diagrams.
We apply the standard method of Feynman
parametrization in evaluating Fig.2.
The result for each diagram turns out to be,
\bea
   (1a.) &=& \frac{1}{( 8 \pi^2 )^2} \frac{1}{24}
           \ln^4 \frac{s}{\lambda^2} \label{i1a}\ ,\\
   (1b.) &=& \frac{1}{( 8 \pi^2 )^2} \frac{1}{12}
           \ln^4 \frac{s}{\lambda^2} \label{i1b}\ ,\\
   (1c.) &=& \frac{1}{( 8 \pi^2 )^2} \left[
                \frac{1}{8} \ln^4 \frac{s}{M^2} - 
               \frac{1}{3} \ln^3 \frac{s}{M^2} \ln \frac{s}{\lambda^2}
             + \frac{1}{4} \ln^2 \frac{s}{M^2} \ln^2 \frac{s}{\lambda^2}
             \right] \label{i1c}\ ,\\
   (1d.) &=& (1f.) = \frac{1}{( 8 \pi^2 )^2} \frac{1}{24}
           \ln^4 \frac{s}{M^2} \label{i1d}\ ,\\
   (1e.) &=& 2 \times (2a.) = 2 \times (2b.) = \frac{1}{( 8 \pi^2 )^2} \left[
                - \frac{1}{12} \ln^4 \frac{s}{M^2} + 
               \frac{1}{6} \ln^3 \frac{s}{M^2} \ln \frac{s}{\lambda^2}
                         \right] \label{i1e}\ ,\\
   (1g.) &=& 2 \times (2c.) = 2 \times (2d.) = \frac{1}{( 8 \pi^2 )^2}
          \frac{1}{12} \ln^4 \frac{s}{M^2} \label{i1g}\ .
\eea
A comment is in order concerning the above results.
The diagrams Fig.1d and 1f (Fig.2c and 2d) lead to the same
result Eq.(\ref{i1d})(Eq.(\ref{i1g})).
As already pointed out by Melles~\cite{M2}, the physical reason
is that the virtuality of the momentum
circulating the loop is determined
by the singularity of the most external propagator
in the diagram in order to produce the DL leading
contribution.
Therefore the singularity from the photon which propagates inside the
$W$ and/or $Z$ loop in Figs.1d and 2c is
already regulated by the $W$ and/or $Z$ mass.

By combining the group factors and the loop integrals,
we find that all non-exponentiating terms are canceled out completely
and obtain the two loop result which is the second
term of the expansion of the exponentiated Sudakov form factor.
\bean
    \Gamma^{(2)} &=&  1 - 
           \frac{1}{16 \pi^2} \left( g^2 C_2 (R) + g'^2 Y^2 - e^2 Q^2 \right)
              \ln^2 \frac{s}{M^2} - \frac{1}{16 \pi^2} e^2 Q^2
                  \ln^2 \frac{s}{\lambda^2} \ , \\
           && + \frac{1}{2!}
            \left[\frac{1}{16 \pi^2} 
       \left( g^2 C_2 (R) + g'^2 Y^2 - e^2 Q^2 \right)
              \ln^2 \frac{s}{M^2} + \frac{1}{16 \pi^2} e^2 Q^2
                  \ln^2 \frac{s}{\lambda^2}\right]^2 \ .
\eean   
The cancellation of the non-exponentiating terms from each diagram
occurs as follows.
The terms coming from the SU(2)$\otimes$U(1) mixing effect
are canceled by the fact that the accompanying integrals 
turn out to be the same.
The mechanism of cancellation of other terms are the same as in QCD. 

\section{Summary}
We have considered the electroweak form factor at two loop
level in the DL approximation.
We have used the standard Feynman gauge.
Our results have shown the exponentiation of the electroweak Sudakov
form factor at two loop level.
The cancellation of the non-exponentiating contributions
is never trivial.
The typical aspects of electroweak theory, the non-abelian nature of 
SU(2) and the mixing effect between SU(2) and U(1),
produce a new situation compared to QCD. 
We have shown that there appear non-exponentiating terms
not only in the crossed ladder diagrams
and diagrams with triple gauge boson coupling 
but also in the ladder diagrams.
The reason is that the propagation of the electroweak bosons
has the contribution
which is not proportional to the Casimir operator.
However these new non-exponentiating terms are canceled out each other
by the dynamical reason that the virtuality of the momentum
circulating the loop is determined
by the singularity of the most external propagator
to the DL accuracy. 
The cancellation of the non-exponentiating terms coming from
the non-abelian nature of SU(2) occurs in the same way as in QCD.
  
\section*{Acknowledgment}

We would like to thank old people for valuable
suggestions.
The work of M. H. was supported in part by the Monbusho Grant-in-Aid
for Scientific Research No.11005244.
The work of H. K. was supported in part by the Monbusho Grant-in-Aid
for Scientific Research No.10000504.


\end{document}